\documentclass[atoms,conferenceproceedings,submit,moreauthors]{Definitions/mdpi}
\renewcommand{\linenumbers}{}
\usepackage{epstopdf}
\usepackage{placeins}
\firstpage{1}
\pubvolume{1}
\issuenum{1}
\articlenumber{0}
\pubyear{2026}
\copyrightyear{2026}
\datereceived{ }
\daterevised{ }
\dateaccepted{ }
\datepublished{ }
\newcommand{\TeIII}{Te\,\textsc{iii}}
\newcommand{\TeIV}{Te\,\textsc{iv}}
\newcommand{\SrII}{Sr\,\textsc{ii}}
\Title{HFR distorted-wave photoionization of Te III: a preliminary benchmark for large-scale atomic data production to model nebular-phase kilonovae}
\Author{J. Deprince $^{1,2}$, L. Maison $^{1}$, C. Guillaume $^{1}$ and E. Bokamba Motoumba $^{3,4}$}
\AuthorNames{J. Deprince, L. Maison, C. Guillaume and E. Bokamba Motoumba}
\address{%
$^{1}$ \quad Atomic Physics and Astrophysics, Universit\'e de Mons, B-7000 Mons, Belgium\\
$^{2}$ \quad Astronomy and Astrophysics Institute, Universit\'e Libre de Bruxelles, B-1050 Brussels, Belgium\\
$^{3}$ \quad Facult\'e des Sciences et Techniques, Universit\'e Marien Ngouabi, Brazzaville P.O. Box 69, Congo\\
$^{4}$ \quad UMR 8262 CNRS, Laboratoire d’\'etude de l’Univers et des Ph\'enom\`enes eXtr\^emes—LUX, Observatoire de Paris-Meudon, PSL, Sorbonne Universit\'e, F-92195 Meudon, France
}
\abstract{The shortage of photoionization cross sections for heavy elements remains an obstacle to non-local thermodynamic equilibrium  (NLTE) modeling of nebular-phase kilonova ejecta, where hydrogenic prescriptions have often been adopted in the absence of ion-specific calculations. We present a first preliminary benchmark of direct photoionization cross sections for doubly ionized tellurium, obtained using the distorted-wave (DW) approximation within the pseudo-relativistic Hartree--Fock (HFR) method implemented in the Cowan codes, in which a semi-empirical adjustment based on the observed energy levels is performed. For the photoionization from the ground level of Te III to the ground level of Te IV, comparison with Dirac R-matrix data shows encouraging agreement in the magnitude and overall decrease of the non-resonant background in the lower photon-energy region, approximately below 4 Ry (autoionizing resonances are absent from the present direct treatment). These preliminary results support HFR distorted-wave calculations as a promising and computationally tractable approach for large-scale production of photoionization cross sections, with lanthanides and actinides as longer-term targets, in light of their importance to kilonova modeling and of the severe lack of atomic data for these species.}
\keyword{atomic data; tellurium; photoionization; distorted wave; Hartree--Fock; kilonovae; non-LTE}

\begin{document}

\section{Introduction}
Kilonova spectra provide a means of studying heavy-element production by rapid neutron capture in neutron-star mergers. Their interpretation requires atomic data for a large number of species, including heavy ions with complex electronic structures. Tellurium is of particular interest following the proposed identification of [\TeIII] emission near $2.1\,\mu$m in AT2017gfo and AT2023vfi \cite{hotokezaka2023,levan,mul2024,gillanders}.

In the dilute nebular-phase kilonova ejecta, non-local thermodynamic equilibrium (NLTE) modeling requires rates for radiative and collisional processes rather than populations determined solely by Boltzmann and Saha relations. Photoionization is one of these processes, but suitable cross sections remain scarce for many $r$-process species, particularly for lanthanides and actinides. In this context, Pognan et al. explicitly report adopting hydrogenic cross sections for their radiative-transfer simulations of kilonovae because of this shortage \cite{pognan2022}. Thus, there is a practical need for reliable calculations of actual photoionization cross sections for a large set of heavy elements.

Close-coupling R-matrix calculations provide a detailed description of photoionization, including continuum-channel coupling and resonance structure. Their computational demands increase with the complexity of the target, making a comprehensive survey of heavy ions particularly challenging. The distorted-wave (DW) approximation offers a more tractable description of the direct process. Combined with a multiconfiguration bound-state model, it can supply transition amplitudes and continuum wavefunctions to compute photoionization cross sections at a lower computational cost than close-coupling calculations, while offering a detailed description going well beyond a generic hydrogenic prescription.

This strategy follows the same motivation as our benchmark of approximate electron-impact excitation calculations for \SrII\ \cite{sr}: assess accessible atomic methods against detailed reference data in support of calculations across many species. This work presents a first preliminary photoionization benchmark discussed at the conference ASOS15. It focuses on a well-defined partial channel, from the ground level of \TeIII\ to the ground level of \TeIV, for which Dirac R-matrix photoionization cross sections are available \cite{mul2026}.

\section{Atomic models and method}
\subsection{Multiconfiguration HFR atomic structure}
The pseudo-relativistic Hartree--Fock (HFR) method implemented in the Cowan codes \cite{cowan} uses configuration-dependent radial orbitals and multiconfiguration expansions of the atomic states. Spin--orbit, mass--velocity and Darwin contributions are included, together with configuration interaction within the chosen expansion. Radial parameters can be semi-empirically adjusted through a least-squares fit to observed energy levels. The HFR method is described in detail in \cite{cowan}, while a summary can be found in our previous works (e.g., \cite{opacity}).

The \TeIII\ multiconfiguration model that we considered contains 25 even-parity and 25 odd-parity configurations, listed in Table~\ref{tab:config} with the [Kr]4d$^{10}$ core omitted. The structure calculation uses 69 observed levels, comprising 14 even and 55 odd levels, from the spectroscopic information compiled in NIST and the analysis of Tauheed and Naz \cite{nist,tauheed}. Table~\ref{tab:levels} illustrates the HFR energy levels before and after adjustment for the lowest excited levels of the ground configuration. The fitted values reproduce the observed $^3$P$_1$, $^3$P$_2$ and $^1$D$_2$ excitation energies to within about $0.3\%$. The fitted parameters for the ground configuration are listed in Table~\ref{tab:fit_parameters}.

\begin{table}[ht]
\centering\small
\caption{Valence configurations of the \TeIII\ model. Each brace denotes all listed choices; the common [Kr]4d$^{10}$ core is omitted.}
\label{tab:config}
\begin{tabularx}{\linewidth}{@{}XX@{}}\toprule
Even parity (25 configurations) & Odd parity (25 configurations) \\\midrule
5s$^2$5p$^2$; 5s$^2$5p\{6p,7p,8p,4f,5f,6f\} & 5s$^2$5p\{6s,7s,8s,5d,6d,7d,5g\} \\
5s5p$^2$\{6s,7s,8s,5d,6d,7d,5g\} & 5s5p$^3$; 5s5p$^2$\{6p,7p,8p,4f,5f,6f\} \\
5s5p6s\{6p,7p,4f,5f\} & 5s5p6s$^2$; 5s5p6s\{7s,5d,6d\} \\
5s5p6p7s; 5s5p5d\{6p,5f\} & 5s5p6p$^2$; 5s5p4f$^2$ \\
5s5p4f\{7s,8s,5d,6d\} & 5s5p4f\{6p,5f\}; 5s5p7s$^2$ \\
 & 5s5p5d$^2$; 5s5p5d7s \\\bottomrule
\end{tabularx}
\end{table}

\begin{table}[ht]
\centering\small
\caption{Excitation energies (cm$^{-1}$) for the lowest excited levels of 5s$^2$5p$^2$ in \TeIII. The ground level is $^3$P$_0$.}
\label{tab:levels}
\begin{tabular}{@{}lccr@{}}\toprule
Level & Observed & HFR & HFR+fit \\\midrule
$^3$P$_1$ & 4757.2 & 4109.1 & 4769.0 \\
$^3$P$_2$ & 8167.0 & 7166.4 & 8149.9 \\
$^1$D$_2$ & 17359.8 & 15500.3 & 17364.6 \\
$^1$S$_0$ & 30398.3 & 30424.2 & 30398.3 \\\bottomrule
\end{tabular}
\end{table}

\begin{table}[ht]
	\centering
	\small
	\caption{Fitted parameters for the ground configuration
		5s$^2$5p$^2$ of \TeIII. The last column gives the ratio of the
		fitted value to its ab initio HFR value.}
	\label{tab:fit_parameters}
	\begin{tabular}{@{}lrr@{}}
		\toprule
		Parameter & Fitted value (cm$^{-1}$) & Ratio \\
		\midrule
		$E_{\mathrm{av}}$ & 15147 & -- \\
		$F^2$(5p,5p)      & 31155 & 0.68 \\
		$\alpha$          &   275 & -- \\
		$\zeta$(5p)       &  5452 & 0.79 \\
		\bottomrule
	\end{tabular}
\end{table}

The \TeIV\ residual-ion model contains 18 configurations, also with a closed 4d$^{10}$ core. The even-parity valence configurations are 5s$^2$\{6s,7s,5d,6d\}, 5s5p$^2$, 5s5p\{6p,4f\}, 5s6p$^2$, 5s5d$^2$, 5s5d6s and 5p$^2$5d. The odd-parity configurations are 5s$^2$\{5p,6p,4f\}, 5p$^3$, 5s5p\{6s,5d\} and 5s5d6p. The reference R-matrix target uses a larger configuration expansion that additionally includes open-4d configurations and retains 100 target levels \cite{mul2026}.

\subsection{Direct photoionization and energy scale}
The outgoing electron is represented by a distorted continuum wave in the residual-ion potential. Electric-dipole matrix elements between the multiconfiguration initial state and the final ion-plus-electron state give the direct photoionization cross section. This description takes into account the interaction of the continuum electron with the ionic potential. However, it does not include the autoionizing resonances arising in the close-coupling calculation.

The photoionization channel considered in this preliminary work is the transition from the ground level of \TeIII\ to the ground level of \TeIV, i.e.
\begin{equation}
 {\rm5s^25p^2\,{}^3P_0}\;(\text{\TeIII})+h\nu
 \longrightarrow {\rm5s^25p\,{}^2P_{1/2}}\;(\text{\TeIV})+e^-.
 \label{eq:channel}
\end{equation}
The photon energy is related to the outgoing-electron kinetic energy $\epsilon$ by
\begin{equation}
 E_\gamma=\chi_0+\epsilon,
 \label{eq:energy}
\end{equation}
where $\chi_0$ denotes the ionization energy for the specific channel. In the present HFR calculation, this energy is approximated by the difference between the average energies of the ground configurations of \TeIII\ and \TeIV, yielding $\chi_0=1.992$ Ry. Mulholland et al.~\cite{mul2026} report a ground-level ionization energy of $2.099$ Ry, while NIST gives a value of $2.046$ Ry \cite{nist}. Our HFR ionization energy is thus in good agreement with the NIST value, underestimating it by about 2.6\%, while the value reported by Mulholland et al. overestimates it by the same amount. The cross sections are plotted against photon energy using the ionization energy adopted in each calculation (without shifting the curves to a common threshold).

\section{Ground-to-ground photoionization channel benchmark}
\begin{figure}[ht!]
\centering
\includegraphics[width=0.60\linewidth]{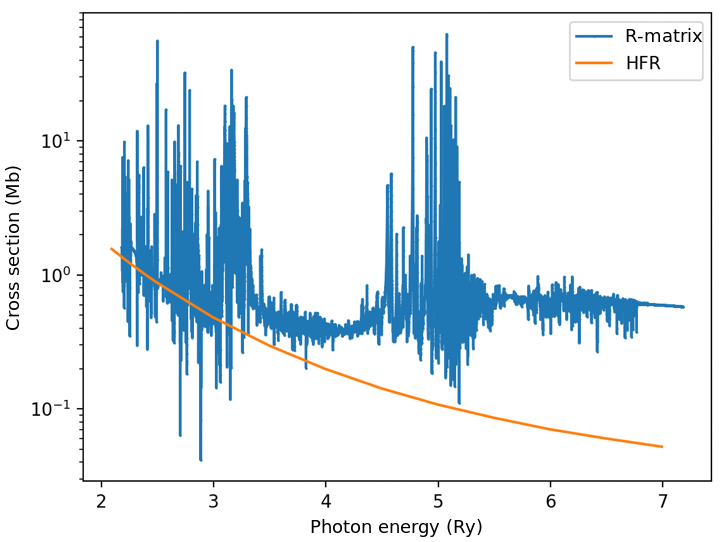}
\caption{Partial photoionization cross section from the \TeIII\ ground level 5s$^2$5p$^2$\,{}$^3$P$_0$ to the \TeIV\ ground level 5s$^2$5p\,{}$^2$P$_{1/2}$. Orange: HFR/DW calculation; blue: R-matrix reference data \cite{mul2026}. The horizontal axis denotes photon energy in Ry and the vertical axis cross section in Mb.}
\label{fig:comparison}
\end{figure}

Figure~\ref{fig:comparison} compares the HFR partial direct photoionization cross section with the corresponding R-matrix data of Mulholland et al. \cite{mul2026}. The HFR/DW result decreases smoothly from $1.5614$ Mb at $\epsilon=0.1$ Ry to $0.0522$ Mb at $\epsilon=5.0$ Ry. The encouraging feature of this comparison is the behavior at lower photon energies. Below approximately $4$ Ry, the HFR/DW direct photoionization cross section follows the overall decline and remains comparable in magnitude to the non-resonant background\footnote{Here, non-resonant background refers to the smooth underlying behavior visible between the resonance structures, approximately corresponding to the direct part of the photoionization cross section.} of the R-matrix cross section, while a discrepancy appears at higher energies. The agreement is qualitative and varies across this interval: the curves progressively diverge as the photon energy increases, with a more pronounced discrepancy above approximately $4$ Ry. This value therefore characterizes the approximate extent of the favorable comparison in Figure~\ref{fig:comparison}, rather than a sharply defined boundary of validity of our HFR/DW approach. Nevertheless, recovering the scale and broad energy dependence of the direct cross section provides a useful first test of the combined
multiconfiguration HFR and DW continuum description.

The numerous peaks in the R-matrix result arise from autoionizing resonance structures absent from the present direct DW treatment. Their absence is an expected feature of the approximation and does not prevent comparison of the direct part of the cross section. The present result is thus encouraging specifically for the direct component and does not
quantify the contribution of the resonances to the photoionization cross section.

At higher photon energies (above 4 Ry), the difference extends beyond the individual resonance peaks: the HFR/DW cross section continues to decrease, whereas the R-matrix background varies more slowly, with the discrepancy approaching an order of magnitude towards the upper end of the considered energy range. The agreement is therefore limited to low energies below approximately 4 Ry. 

Taken together, these results identify the low-energy direct component of the photoionization cross section as the favorable outcome of this first preliminary benchmark and support the use of HFR/DW as a practical approximate approach to producing photoionization data for heavy ions.

\FloatBarrier
\section{Relevance to atomic data production for nebular-phase kilonova modeling}
This preliminary benchmark for the ground-to-ground photoionization of \TeIII\ suggests that the HFR/DW approach offers a practical compromise between computationally demanding R-matrix calculations and simple hydrogenic prescriptions, providing a physically motivated alternative for heavy ions without detailed photoionization data available. The present comparison shows that this tractable treatment can recover the scale and energy dependence of the low-energy direct part of the photoionization cross sections for a heavy ion. Similarly to the approximate approach that we developed for electron-impact excitation collision strengths and which we benchmarked against R-matrix data in the case of \SrII\ \cite{sr}, this is a positive and encouraging result regarding photoionization cross sections for a large-scale effort aimed at filling gaps in atomic data sets required for NLTE nebular-phase kilonova modeling. The methodological advance over a hydrogenic prescription is the explicit multiconfiguration description of the bound states and ion-specific continuum potential, which is found to be able to reproduce the low-energy direct photoionization cross section in the specific case of the ground-to-ground photoionization channel of \TeIII.

Pognan et al. find that photoionization is typically more effective for neutral and singly ionized species in their NLTE kilonova models, whereas non-thermal ionization dominates for doubly ionized species; they attribute this trend to the higher thresholds and the moderate-energy radiation field \cite{pognan2023}. This supports attention to the lower-energy part of photoionization cross sections, although it does not establish a universal photon-energy boundary. In particular, $4$ Ry corresponds to about $54.4$ eV, and the contribution of photons above this energy to the photoionization rate depends on the intensity of the local radiation field at these energies. The importance of the high-energy part of the cross section where a departure is observed between our HFR/DW approach and the reference R-matrix calculation (see Figure~\ref{fig:comparison}) is therefore specific to the radiation field and ionization balance of the ejecta.

The principal motivation for HFR/DW is the combination of atomic specificity and computational tractability. It builds on a framework already used for extensive heavy-element bound-state calculations, including our large-scale atomic data and LTE opacity calculations from Ca to Lr \cite{opacity}. A common approach is attractive for producing data across many ions, especially for lanthanides and actinides, where the need for atomic data for NLTE modeling extends far beyond the species with detailed complex calculations. Tellurium serves here as an initial test with an available close-coupling reference. The favorable low-energy comparison thus provides a concrete basis for this approach, while its scope remains that of a preliminary single-channel benchmark.

\section{Conclusions}
We have presented an initial HFR distorted-wave benchmark for photoionization from the ground level of \TeIII\ to the ground level of \TeIV. The multiconfiguration structure model incorporates observed energy levels through semi-empirical adjustment. The resulting direct cross section shows encouraging agreement with the magnitude and overall decrease of the R-matrix non-resonant background at lower photon energies, approximately below 4 Ry, with the closest agreement in the lower part of that interval. Resonances are absent from the present direct treatment, and noticeable differences develop at higher energies.

These results support HFR/DW as a promising approach to produce direct photoionization cross sections where detailed data are unavailable and hydrogenic prescriptions are otherwise used. Together with our earlier \SrII\ electron-impact excitation study, in which a similar approach was succesfully tested again R-matrix calculations, this preliminary benchmark contributes to the development of a computationally tractable approach for atomic data production for NLTE nebular-phase kilonova modeling. 

Building on this preliminary benchmark, future work will focus on refining the multiconfiguration atomic models and their semi-empirical adjustment to observed energy levels. Particular emphasis will be placed on extending the calculations to multiple photoionization channels, including those leaving the residual ion in excited levels. Summing the contributions of the accessible final channels will provide total direct photoionization cross sections for each initial level considered. This approach could also be extended to include resonant photoionization through separate calculations of photoexcitation to autoionizing states and their subsequent decay. The resonant contributions could then be combined with the direct cross sections in a post-processing step using the independent-processes and isolated-resonance distorted-wave (IPIRDW) approximation, neglecting interference between the direct and resonant pathways \cite{delahaye2013}. A similar treatment of resonant contributions has recently been applied to electron-impact excitation of \TeIII\ in the same context by Ferreira da Silva et al.~\cite{silva2026}.

The broader objective of this work is to supply consistent sets of atomic data for many heavy ions relevant to nebular-phase kilonova modeling, including lanthanides and actinides, as well as to extend our approach to other relevant atomic processes.

\begin{adjustwidth}{-\extralength}{0cm}
\reftitle{References}

\end{adjustwidth}
\end{document}